\documentclass[twocolumn]{aastex63}
\usepackage{graphicx}
\usepackage{epstopdf}
\usepackage{amsmath}

\received{Aug 17, 2020}
\accepted{Oct 28, 2020}
\published{Dec 31, 2020}

\shorttitle{Radiation pressure in obscured quasars}
\shortauthors{Jun et al.}

\def\deg{\ifmmode {^{\circ}}\else {$^\circ$}\fi}
\def\kms{\ifmmode {\rm\,km\,s^{-1}}\else
    ${\rm\,km\,s^{-1}}$\fi}
\def\ergcm2s{\ifmmode {\rm\,erg\,cm^{-2}\,s^{-1}}\else
    ${\rm\,erg\,cm^{-2}\,s^{-1}}$\fi}
\def\ergAcm2s{\ifmmode {\rm\,erg\,cm^{-2}\,s^{-1}\,\AA^{-1}}\else
    ${\rm\,erg\,cm^{-2}\,s^{-1}\,\AA^{-1}}$\fi}
\def\ergs{\ifmmode {\rm\,erg\,s^{-1}}\else
    ${\rm\,erg\,s^{-1}}$\fi}
\def\kmsMpc{\ifmmode {\rm\,km\,s^{-1}\,Mpc^{-1}}\else
    ${\rm\,km\,s^{-1}\,Mpc^{-1}}$\fi}

\begin{document}
\title{The Dust-to-Gas Ratio and the Role of Radiation Pressure in Luminous, Obscured Quasars}
\author{Hyunsung D. Jun}
\affiliation{School of Physics, Korea Institute for Advanced Study, 85 Hoegiro, Dongdaemun-gu, Seoul 02455, Republic of Korea} \footnote{e-mail: hsjun@kias.re.kr}
\author{Roberto J. Assef}
\affiliation{N\'{u}cleo de Astronom\'{i}a de la Facultad de Ingenier\'{i}a y Ciencias, Universidad Diego Portales, Av. Ej\'{e}rcito Libertador 441, Santiago, Chile}
\author{Christopher M. Carroll}
\affiliation{Department of Physics \& Astronomy, Dartmouth College, Hanover, NH 03755, USA}
\author{Ryan C. Hickox}
\affiliation{Department of Physics \& Astronomy, Dartmouth College, Hanover, NH 03755, USA}
\author{Yonghwi Kim}
\affiliation{School of Physics, Korea Institute for Advanced Study, 85 Hoegiro, Dongdaemun-gu, Seoul 02455, Republic of Korea}
\author{Jaehyun Lee}
\affiliation{School of Physics, Korea Institute for Advanced Study, 85 Hoegiro, Dongdaemun-gu, Seoul 02455, Republic of Korea}
\author{Claudio Ricci}
\affiliation{N\'{u}cleo de Astronom\'{i}a de la Facultad de Ingenier\'{i}a y Ciencias, Universidad Diego Portales, Av. Ej\'{e}rcito Libertador 441, Santiago, Chile}
\affiliation{Kavli Institute for Astronomy and Astrophysics, Peking University, Beijing 100871, People’s Republic of China}
\author{Daniel Stern}
\affiliation{Jet Propulsion Laboratory, California Institute of Technology, 4800 Oak Grove Drive, Pasadena, CA 91109, USA}

\begin{abstract}
The absence of high-Eddington-ratio, obscured active galactic nuclei (AGNs) in local ($z\lesssim0.1$) samples of moderate-luminosity AGNs has generally been explained to result from radiation pressure on the dusty gas governing the level of nuclear ($\lesssim10$\,pc) obscuration. However, very high accretion rates are routinely reported among obscured quasars at higher luminosities and may require a different feedback mechanism. We compile constraints on obscuration and Eddington ratio for samples of X-ray, optical, infrared, and submillimeter selected AGNs at quasar luminosities. Whereas moderate-luminosity, obscured AGNs in the local universe have a range of lower Eddington ratios ($f_{\rm Edd} \sim 0.001–0.1$), the most luminous ($L_{\rm bol} \gtrsim 10^{46} \ergs$) IR/submillimeter-bright, obscured quasars out to $z\sim3$ commonly have very high Eddington ratios ($f_{\rm Edd} \sim 0.1–1$). This apparent lack of radiation-pressure feedback in luminous, obscured quasars is likely coupled with AGN timescales, such that a higher fraction of luminous, obscured quasars are seen because of the short timescale for which quasars are most luminous. When adopting quasar evolutionary scenarios, extended ($\sim10^{2-3}$\,pc) obscuration may work together with the shorter timescales to explain the observed fraction of obscured, luminous quasars, while outflows driven by radiation pressure will slowly clear this material over the AGN lifetime. 
\end{abstract}


\section{Introduction}
Accretion rate and obscuration of active galactic nuclei (AGNs) are two fundamental parameters that explain their diverse observed properties (e.g., \citealt{She14}; \citealt{Hic18}). According to AGN evolutionary scenarios (e.g., \citealt{Hop08}; \citealt{Hic09}), matter falling onto the black hole (BH) will trigger AGN activity at various strengths depending on the environment, which in turn will heat and sweep any potential obscuring circumnuclear material. Thus, feeding onto the BH and feedback from the AGN are closely related on the obscuration--accretion rate plane. 

\citet{Ric17c} (hereafter R17c) recently reported on a study of the relationship between obscuration and accretion rate in a large, relatively unbiased, and complete sample of local AGNs. Specifically, they investigated 836 AGNs with a median redshift of $\langle z \rangle = 0.037$ selected by the hard X-ray (14--195~keV) {\it Swift} Burst Alert Telescope (BAT, \citealt{Geh04}; \citealt{Bar05}; \citealt{Kri13}) all-sky survey (\citealt{Bau13}; \citealt{Kos17}; \citealt{Oh18}), which is sensitive to sources with column densities up to $N_H \approx 10^{24}\, {\rm cm}^{-2}.$  Approximately one-half of the sources had robust measurements of column densities, intrinsic X-ray luminosities, and black hole masses, from which R17c was able to show that while unobscured AGNs are seen with Eddington fractions up to the Eddington limit, very few local, obscured AGNs are found with Eddington fractions above approximately 10\%. This strengthened earlier results based on smaller samples (e.g., \citealt{Fab09}) and was interpreted as evidence for radiation-pressure-driven AGN feedback (e.g., \citealt{Kin03}; \citealt{Mur05}) clearing the immediate BH environment of dusty gas (e.g., \citealt{Fab06,Fab08}). For dusty gas (neutral or partially ionized), the effective cross section between matter and radiation ($\sigma_{\rm dust}$) becomes larger than that between electrons and radiation for ionized gas ($\sigma_{\rm T}$, for Thompson scattering), due to absorption of radiation by dust. This is given by the Eddington ratio for ionized gas, 
\begin{eqnarray}\begin{aligned}
f_{\rm Edd}=L_{\rm bol}/L_{\rm Edd}\\
L_{\rm Edd}=4\pi GM_{\rm BH}&m_{\rm p}c/\sigma_{\rm T}
\end{aligned}\end{eqnarray}
where $L_{\rm bol}$ is the bolometric luminosity, $L_{\rm Edd}$ is the Eddington luminosity with $M_{\rm BH}$ the BH mass, and $m_{\rm p}$ is the proton mass. For dusty gas, we use $\sigma_{\rm dust}$ instead of $\sigma_{\rm T}$ in Equation (1), where $f_{\rm Edd}$ is redefined as the effective Eddington ratio ($f_{\rm Edd, dust}=f_{\rm Edd}\sigma_{\rm dust}/\sigma_{\rm T}$). AGNs are strong ionizing sources, but are fully ionized close to their accretion disks (e.g., \citealt{Ost79}; \citealt{Bal01}), though $\gtrsim$pc-scale environment starts to be composed of dusty gas (e.g., \citealt{Kis11}; \citealt{Min19}).

\begin{figure}
\centering
\includegraphics[scale=.95]{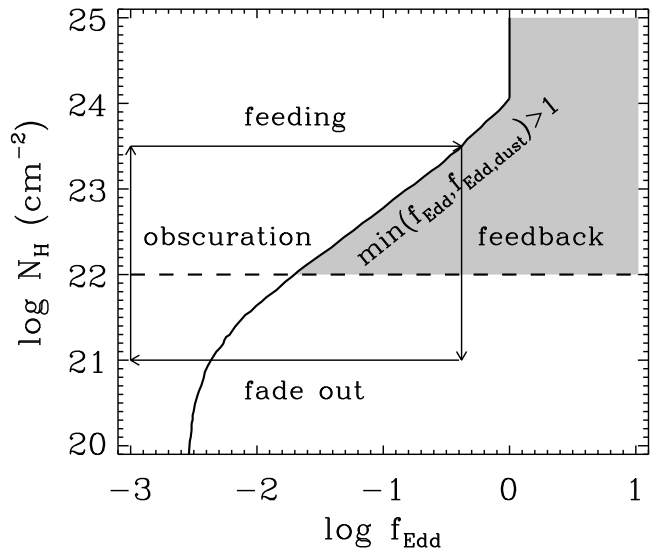}
\caption{The $N_{\rm H}$--$f_{\rm Edd}$ plane showing a schematic evolution of local AGNs probed by the {\it Swift}/BAT X-ray data (R17c). The shaded region indicates the region unoccupied in R17c, which is explained as radiation pressure quickly blowing out the nuclear obscuration composed of dusty gas. In this view, the nuclear obscuration is followed by feeding of gas onto supermassive BHs, where the obscuration is suppressed by radiation-pressure-driven feedback until the accretion starves and the AGN fades out.} 
\end{figure}

To visualize how the radiation pressure would blow away the nuclear material, we plot $f_{\rm Edd}$ where $f_{\rm Edd, dust}=1$ as a function of $N_{\rm H}$ in Figure 1, adopted from \citet{Fab09}.\footnote{We note that at $N_{\rm H}\gtrsim10^{24}\, \rm cm^{-2}$, $f_{\rm Edd, dust}<f_{\rm Edd}$, or the radiation pressure on dusty gas is weaker than on ionized gas. Only because of the relative lack of super-Eddington gas accretion, we restrict the $f_{\rm Edd}(f_{\rm Edd, dust}=1)$ curve to $f_{\rm Edd}<1$ throughout.} When AGNs are obscured beyond the typical $N_{\rm H}\sim10^{22}\, \rm cm^{-2}$ value for galaxies (e.g., \citealt{Buc17}; \citealt{Hic18}), $f_{\rm Edd}(f_{\rm Edd, dust}=1)$ is on the order of 0.01--0.1, and the accretion becomes super-Eddington (shaded region) even if $f_{\rm Edd}(f_{\rm Edd, dust}=1)<f_{\rm Edd}<1$. This makes it easier for radiation pressure to blow out the dusty nuclear gas on the order of 1--10 pc (e.g., \citealt{Jaf04}; \citealt{Ram17}; \citealt{Ich19}) and matches well with the observations of R17c reporting a lack of AGNs with $f_{\rm Edd, dust}>1$, perhaps favoring an episodic nuclear obscuration and blowout governed by radiation pressure. 

The largest limitation for the $\it Swift$/BAT survey, which is relatively unbiased and complete for local AGNs, is that its shallow sensitivity misses luminous quasars in the distant universe. Here, we explore high-luminosity/redshift samples of optical quasars (e.g., \citealt{Sch10}), optical--IR red quasars with large color excess, where $E(B-V)\lesssim1.5$ mag (e.g., \citealt{Gli07}; \citealt{Ban12}; \citealt{Ros15}; \citealt{Ham17}), dust-obscured galaxies (DOGs, \citealt{Dey08}; Hot DOGs, \citealt{Eis12}), and submillimeter galaxies (SMGs, \citealt{Bla02}), where the latter two are likely subsets and distant analogs of local ultraluminous infrared galaxies (ULIRGs, $\log L_{\rm IR}>10^{12} L_{\odot}$; \citealt{San88}) and their higher-luminosity cousins (e.g., hyLIRGs, \citealt{San96}; ELIRGs, \citealt{Tsa15}). We also add the highest-obscuration Compton-thick AGNs, that is, AGNs with X-ray-obscuring column densities of $N_{\rm H}\gtrsim10^{24}\, \rm cm^{-2}$, observed with $\it NuSTAR$ \citep{Har13}.

\begingroup
\begin{deluxetable*}{ccccccc}
\tablecolumns{7}
\tabletypesize{\scriptsize}
\tablecaption{AGN samples}
\tablehead{
\colhead{Name} & \colhead{Sample} & \colhead{Selection} & \colhead{Obscuration} & \colhead{$z$} & \colhead{$\log L_{\rm bol}\,(\ergs)$} & \colhead{N}}
\startdata
B15b; B16 & Compton-thick & Hard X-ray & $N_{\rm H}$ & 0.001--0.051 & 42.5--45.8 & 16\\
R17c & {\it Swift}/BAT & Hard X-ray & $N_{\rm H}$, $E(B-V)_{\rm nl}$ & 0.00--0.27 & 40.8--46.9 & 366\\
Y09; M17; V18b & Type 1 & Optical & $N_{\rm H}$, $E(B-V)_{\rm cont}$ & 0.15--4.26 & 44.8--48.7 & 174\\
J13 & Type 1 & Optical & $E(B-V)_{\rm cont}$ & 0.14--4.13 & 45.7--48.2 & 14,531\\ 
L16; L17; G17& Red & Optical--IR & $N_{\rm H}$, $E(B-V)_{\rm cont}$ & 0.14--2.48 & 45.2--46.9 & 12\\ 
U12; K18 & Red & Optical--IR & $E(B-V)_{\rm cont}$ & 0.29--0.96 & 45.8--47.1 & 23\\ 
G18 & Extremely red & Optical--IR & $N_{\rm H}$ & 2.32 & 47.5 & 1\\ 
P19 & Extremely red & Optical--IR & $E(B-V)_{\rm cont}$ & 2.24--2.95 & 46.7--47.8 & 28\\ 
L20 & Heavily reddened & Optical--IR & $N_{\rm H}$ & 2.09--2.66 & 45.7--46.9 & 7\\ 
B12; B13; B15a; T19 & Heavily reddened & Optical--IR & $E(B-V)_{\rm cont}$ & 1.46--2.66 & 46.0--48.6 & 51\\
C16; T20 & DOG & Optical--IR & $N_{\rm H}$ & 1.22--5.22 & 43.8--48.2 & 15\\
S14; A16; R17a; V18a; Z18; A20 & Hot DOG & IR & $N_{\rm H}$ & 1.01--4.60 & 46.2--48.1 & 9\\ 
A15 & Hot DOG & IR & $E(B-V)_{\rm cont}$ & 0.29--4.59 & 45.4--48.9 & 129\\
A08 & ULIRG, SMG & IR/submm  & $N_{\rm H}$ & 0.04--2.05 & 44.6--45.6 & 3\\
\enddata
\tablecomments{AGN samples used in this work. We refer to each reference as having the objects with values or constraints on obscuration or accretion rate, while the original catalog paper is provided in the text (\S2). Here, $N$ denotes the number of objects for each set of references having obscuration and accretion rate information used in \S5, so they could be smaller than the number of sources from the references. The subscripts ``cont'' and ``nl'' under $E(B-V)$ values are derived using the continuum SED and narow-line ratios, respectively. The abbreviated references are \citet{Bri15} (B15b); \citet{Bri16} (B16); \citet{Ric17c} (R17c); \citet{You09} (Y09); \citet{Mar17} (M17); \citet{Vie18} (V18b); \citet{Jun13} (J13); \citet{Lam16,Lam17} (L16, L17); \citet{Gli17a} (G17); \citet{Urr12} (U12); \citet{Kim18} (K18); \citet{Gou18} (G18); \citet{Per19} (P19); \citet{Lan20} (L20); \citet{Ban12,Ban13,Ban15} (B12, B13, B15a); \citet{Tem19} (T19); \citet{Cor16} (C16); \citet{Tob20} (T20); \citet{Ste14} (S14); \citet{Ass16} (A16); \citet{Ric17a} (R17a); \citet{Vit18} (V18a); \citet{Zap18} (Z18); \citet{Ass20} (A20); \citet{Ass15} (A15); \citet{Ale08} (A08).} 
\end{deluxetable*} 
\endgroup 

The latest studies of obscured quasars with large $E(B-V)$ values, through careful analysis to quantify and minimize the effect of obscuration, have reported near-Eddington to Eddington-limited accretion ($f_{\rm Edd}\sim$\,0.1--1, e.g., \citealt{Ale08}; \citealt{Urr12}; \citealt{Kim15}; \citealt{Ass20}; \citealt{Jun20}). Furthermore, \citet{Gli17b} and \citet{Lan20} find many obscured quasars with $f_{\rm Edd, dust}>1$ at high $N_{\rm H}$. These observations suggest that radiation pressure on dusty gas is effective, but is potentially less effective for obscured, luminous quasars since the length of time that luminous quasars are active is shorter than the length of time that less-luminous AGN are active (e.g., \citealt{Hop05, Hop06}). Alternatively, luminous, obscured quasars are thought to be observed in a short phase in which they are blowing out the material through outflows stronger at higher luminosities (e.g., \citealt{Lam17}; \citealt{Per19}; \citealt{Tem19}; \citealt{Jun20}), perhaps requiring a different nuclear or galactic environment from less-luminous, obscured AGNs. Hence, there is growing interest in which AGN property drives radiation-pressure feedback, and in which temporal and spatial scales is it effective.

In this work, we attempt to constrain the $N_{\rm H}$--$f_{\rm Edd}$ and $E(B-V)$--$f_{\rm Edd}$ planes for quasars from multiwavelength AGN samples (\S2) and through a consistent method to estimate $N_{\rm H}$, $E(B-V)$ (\S3), and $f_{\rm Edd}$ values (\S4). We present (\S5) and discuss (\S6) the $N_{\rm H}$--$f_{\rm Edd}$ and $E(B-V)$--$f_{\rm Edd}$ distributions for quasars in terms of various feedback mechanisms. Throughout, including the luminosities from the literature, we use a flat $\Lambda$CDM cosmology with $H_{0}=\mathrm{70\,km\,s^{-1}\,Mpc^{-1}}$, $\Omega_{m}=0.3$, and $\Omega_{\Lambda}=0.7$.

\section{The sample}
Probing the distribution of $N_{\rm H}$--$f_{\rm Edd}$ and $E(B-V)$--$f_{\rm Edd}$ values from a statistically complete AGN sample is complicated for several reasons. AGNs radiate across almost the entire electromagnetic spectrum, but show a wide range of spectral energy distributions (SEDs) due to physical processes governing the radiation, host galaxy contamination, and obscuration on various scales around the accreting BHs (e.g., \citealt{Lan17}; \citealt{Hic18}). Therefore, we found it beneficial to compile quasar samples selected at various wavelengths over a wide range of luminosity and redshift. Still, we chose to add only the data from the literature that meaningfully increase the sample size for a given wavelength selection.

\begin{figure*}
\centering
\includegraphics[scale=.95]{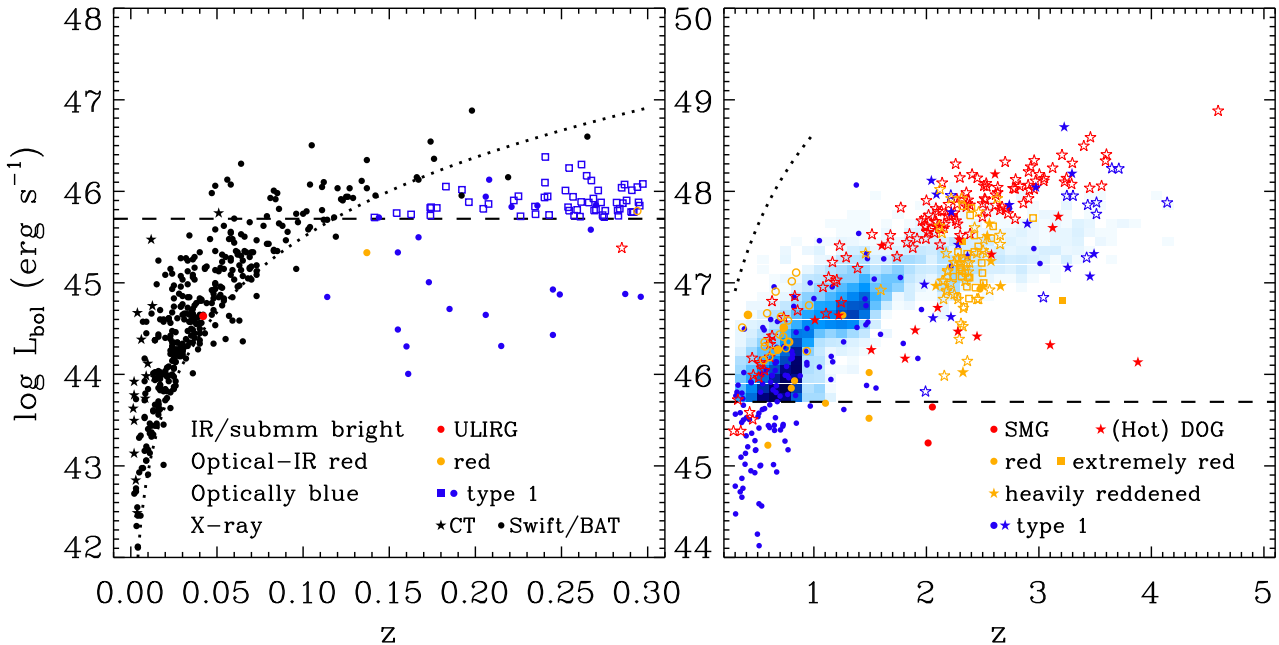}
\caption{Bolometric luminosity ($L_{\rm bol}$) as a function of redshift (left: $z<0.3$, right: $z>0.3$) for AGNs having $f_{\rm Edd}$ and measurements of either $N_{\rm H}$ (filled symbols) or $E(B-V)$ (open symbols, plotted together if they have both $N_{\rm H}$ and $E(B-V)_{\rm cont}$). The samples plotted are ULIRGs/SMGs (A08, red circles), Hot DOGs (S14; A15; A16; R17a; V18a; Z18; A20, red stars), optical--IR red AGN samples named red quasars (U12; L16, L17; G17; K18, yellow circles), extremely red quasars (P19, yellow squares), heavily reddened quasars (B12; B13; B15a; T19; L20, yellow stars), Compton-thick AGNs (B15b; B16, black stars), $\it Swift$/BAT AGNs (R17c, black circles), and optically selected SDSS type 1 quasars (J13, blue squares for $z<0.3$, density plot for $z>0.3$ (due to a large sample), matched with Y09 in blue circles; WISSH quasars from M17 and V18b as blue stars). The $L_{\rm bol} \ge 10^{45.7}\ergs$ boundary is marked (dashed line), and a constant 14--195\,keV flux of $10^{-11} \ergs \rm cm^{-2}$ with bolometric correction applied (\S4), roughly denotes the detection limit of {\it Swift}/BAT X-ray data (dotted line, drawn up to $z=1$).}
\end{figure*}

In Figure 2 we plot the AGN samples from X-ray, optically blue, optical--IR red, and IR/submillimeter-bright populations, also summarized in Table 1. At $z<0.3$, the $\it Swift$/BAT AGN in R17c (with $N_{\rm H}$ from \citealt{Ric17d}; $E(B-V)$ from \citealt{Kos17}) has the advantage of minimal obscuration bias from the hard X-ray selection and covers a wide range of luminosities, reaching down to low-luminosity AGNs and up to quasar luminosities ($10^{41}\lesssim L_{\rm bol} \lesssim 10^{47} \ergs$). However, the $\it Swift$/BAT survey lacks the sensitivity to probe the distant or luminous AGN populations marked in Figure 2. We complemented the highest obscurations using Compton-thick ($N_{\rm H} \gtrsim 10^{24}\, \rm cm^{-2}$) AGNs observed with $\it NuSTAR$ (B15b; B16), and the higher luminosities/redshifts from optical Sloan Digital Sky Survey (SDSS) quasars (e.g., \citealt{Sch10}; $N_{\rm H}$ from Y09; $E(B-V)$ from J13), red quasars (L16, originally from \citealt{Gli12}; K18, averaged between \citealt{Gli07} and \citealt{Urr09}), and quasars from ULIRGs (broad-line\footnote{Throughout, we classify narrow-line AGNs to be type $\ge1.8$ (weak broad H$\alpha$ and H$\beta$, [\ion{O}{3}]/H$\beta>3$, \citealt{Win92}), and broad-line AGNs to be type $\le1.5$ (comparable broad-to-narrow H$\beta$ and [\ion{O}{3}]/H$\beta<3$). When the AGN types are not specified, we follow the visual classifications from the literature.} ULIRGs in A08, with $N_{\rm H}$ from \citealt{Sev01}). 

We further include $z>0.3$ quasars to search for luminous quasars, adding type 1 quasars from the SDSS (i.e., WISSH quasars, $N_{\rm H}$ from M17 and $E(B-V)$ from V18b), and a variety of quasars with red optical-to-infrared colors, that is, heavily reddened quasars (B12; B13, B15a; T19), red quasars (U12; G17; L17; K18), extremely red quasars (P19 with $N_{\rm H}$ from G18 and $E(B-V)$ from \citealt{Ham17}), and DOGs (C16; T20). Hot DOGs (S14; A15\footnote{This sample is being updated by P. R. M. Eisenhardt et al. (in preparation), but we simply refer to the numbers from A15 at this time.}; A16; R17a; V18a; Z18; A20), and broad-line (AGN-like) SMGs (A08 with $N_{\rm H}$ from \citealt{Ale05}) were also included, adding part of some samples at $z<0.3$ that extend to $z>0.3$ (A08; Y09; J13; L16; R17c; K18). 

Duplication among the samples was found in Compton-thick AGNs (B15b; B16), heavily reddened quasars (B12; T19) and red quasars (U12; K18), where we used the most recent values, except for those between B15b/B16 and R17c, where we kept both the $N_{\rm H}$ and $f_{\rm Edd}$ estimates as they were based on multiple X-ray observations. The samples based on follow-up studies of SDSS quasars were separated into those with $N_{\rm H}$ (Y09; M17) and those with $E(B-V)$ (J13; V18b), where the $f_{\rm Edd}$ values from signal-to-noise ratio (S/N) $>$20 spectra in \citet{She11} and \citet{Par12} were added. We removed beamed sources (R17c, flagged by \citealt{Kos17} using the blazar catalog from \citealt{Mas15}) for reliable $N_{\rm H}$ and $f_{\rm Edd}$ values (but see also, e.g., \citealt{Bae19}, for estimation of $f_{\rm Edd}$ in radio-bright AGNs). We used line widths corrected for instrumental resolution in estimating $M_{\rm BH}$ (\S4).

\section{Gas and Dust Obscuration}
We compiled $N_{\rm H}$ and $E(B-V)$ values, representing gas and dust obscuration, for the AGN samples. For $N_{\rm H}$, we used the line-of-sight X-ray obscuration from sources with enough X-ray counts to model the spectra ($\gtrsim$40--60, defined by the respective references). Exceptions are obviously large absorption ($N_{\rm H}\ge10^{24}\, \rm cm^{-2}$) constraints in S14, V18a, and A20, where the exposure times are longer than 20 ks but have a relatively smaller number of X-ray counts due to Compton-thick absorption. We add these values to our analysis. The choice of models to fit or estimate the X-ray obscuration varies in the literature: \citet{Mur09} (S14; L16; G17; V18a; Z18), \citet{Bri11} (B15b; A16; B16; C16; A20), hardness-ratio-based $N_{\rm H}$ conversion (L17), (absorbed) power-law fit (Y09; C16; M17; G18; L20), and a combination of models (A08; R17c; T20). Still, when the $N_{\rm H}$ values are compared between various models (e.g., B15b; B16; G17; Z18; L20), they are mostly consistent within the uncertainties (but see also B15b and \citealt{Liu15}, for the limitations of the models at Compton-thick column densities).

For $E(B-V)$, we used the UV/optical--IR continuum SED-based $E(B-V)_{\rm cont}$\footnote{Throughout, $E(B-V)_{\rm cont/bl/nl}$ are those derived from the continuum SED and broad/narrow-line ratios, respectively, and the $E(B-V)_{\rm nl}$ values are only mentioned as lower limits to $E(B-V)_{\rm cont}$. We used the Milky Way extinction curve with total-to-selective extinction of 3.1 when transforming extinction to $E(B-V)$.} from the literature. Lower limits in $E(B-V)$ were given to the P19 data from \citet{Ham17} because of likely underestimation using a narrow range of wavelengths to determine $E(B-V)$. For optical quasars, we determined the rest-frame $>0.3\mu$m power-law continuum slope ${\alpha}$ following $F_{\nu} \propto \nu^{\alpha}$, fit to the photometric SED. We assumed an intrinsic slope of ${\alpha}=0.1\pm0.2$ from the most blue (hot dust-poor, $\sim3\sigma$ outliers) quasars in J13, consistent with accretion disk models and polarized observations of quasar SEDs ($\alpha \approx 1/3$, e.g., \citealt{Sha73}; \citealt{Kis08}). We limited the sample to quasars with at least three SDSS optical or UKIDSS near-IR \citep{Law07} photometric detections at rest-frame $0.3-1\mu$m and rest-frame near-IR detections at up to at least 2.3$\mu$m to decompose the SED into the power-law continuum and dust emission (see J13 for details). We converted $\alpha$ into $E(B-V)$ by reddening the intrinsic slope using a Milky Way extinction curve at $0.3-1\mu$m to match the observed value of $\alpha$, while fixing $E(B-V)=0$ when $\alpha>0.1$. We checked if the $E(B-V)$ estimates from J13 are consistent with the literature by comparing the values cross-matched with 17 sources in V18b and 277 sources in C. Carroll et al. (submitted) at $E(B-V)<0.5$ mag. The J13 $E(B-V)$ values have a median offset and scatter of $0.05\pm0.05$ and $0.04\pm0.06$ mag, respectively, consistent within the uncertainties. We adopt the J13 values for the cross-matched sources (Y09; M17; V18b).

The $E(B-V)_{\rm cont}$ values can suffer from host galaxy contamination in the rest-frame optical/near-IR. We limited the samples with the SEDs decomposed into an AGN and a host galaxy (U12; A15; L17) to $L_{\rm bol} \ge 10^{45.2} \ergs$ for the decomposition to contain a sufficient AGN contribution, and the samples without an SED decomposition (the remaining samples with $E(B-V)_{\rm cont}$ in Table 1) to $L_{\rm bol} \ge 10^{45.7} \ergs$, to minimize host galaxy contamination. Above the luminosity limits, the average host contamination at 5100\AA\ drops below 50\% and 10\%, respectively, for type 1 quasars \citep{She11} and is consistent with the growing AGN contribution to the observed SEDs for red quasars at higher $L_{\rm bol}$ (L17). The $L_{\rm bol} \ge 10^{45.7} \ergs$ limit corresponds to $L_{\rm bol} =10^{12.1} L_{\odot}$, selecting ULIRG luminosities for IR-bright AGNs (e.g., \citealt{Fan16}; \citealt{Tob17}). The majority of the hard X-ray selected AGNs from R17c are less luminous than the $L_{\rm bol}$ limits for $E(B-V)_{\rm cont}$, but instead have robust measurements of $N_{\rm H}$ from their hard X-ray spectra.

The nuclear dust-to-gas ratio traced by the $\log E(B-V)_{\rm cont/bl}/N_{\rm H}$ (mag cm$^{2}$) values in Figure 3 is constant if the gas and dust obscuration are proportional (e.g., \citealt{Usm14}). The values are overall smaller than the Galactic value ($-$21.8, e.g., \citealt{Boh78}),\footnote{One of the reasons for the offset may be that the majority of the AGNs have an excess of dust-free gas within the sublimation radius (e.g., \citealt{Ris02, Ris07}; \citealt{Mai10}; \citealt{Bur16}; \citealt{Ich19}), but we focus here on the overall value when including the more luminous AGNs.} with reported average values ranging between $-$22.8 \citep{Mai01} and $-$22.3 (L20).  L20 find relatively higher $E(B-V)/N_{\rm H}$ values for a sample of heavily reddened broad-line quasars at high luminosity, but there are similarly luminous quasars with relatively smaller $E(B-V)/N_{\rm H}$ values (i.e., the Hot DOGs or some optical--IR red quasars in Figure 3). Apart from the type 1 AGNs where the large scatter in $E(B-V)/N_{\rm H}$ could in part arise from the uncertainty constraining the lowest values in either quantity, we find the mean and scatter of $\log E(B-V)/N_{\rm H} \,(\rm mag\,\,cm^{2})=-22.77 \pm 0.51$ (observed) or $\pm 0.41$\footnote{We refer to the intrinsic scatter of the quantity $x=\log E(B-V)/N_{\rm H}$, $\sigma_{\rm int}$, as the observed scatter with measurement error $\Delta x$ subtracted in quadrature, that is, $\sigma_{\rm int}^{2}=\Sigma_{i=1}^{n}\{(x_{i}+22.77)^{2}-\Delta x_{i}^{2}\}/(n-1)$. The errors on $E(B-V)$ values are missing for the L16 and G17 samples, but the intrinsic scatter of  $\log E(B-V)/N_{\rm H}$ decreases by only 0.01 dex if we assign the mean error of $\Delta E(B-V)=0.12$ mag from the L17 sample used here.} (intrinsic) from 31 obscured AGNs (type 2 AGNs, optical--IR red quasars, and Hot DOGs) without upper/lower limits in Figure 3, spanning absorption-corrected $L_{\rm 2-10 keV}=10^{42.4-45.6}\ergs$. The ratios are close to the \citet{Mai01} value, but are highly scattered for any combination of AGN type over the luminosity probed, complicating a simple correspondence between dust and gas. We thus refer to both $E(B-V)$ and $N_{\rm H}$ when selecting AGNs with dusty gas, using a conversion of $\log E(B-V)/N_{\rm H}=-$22.8.

\begin{figure}
\centering
\includegraphics[scale=0.95]{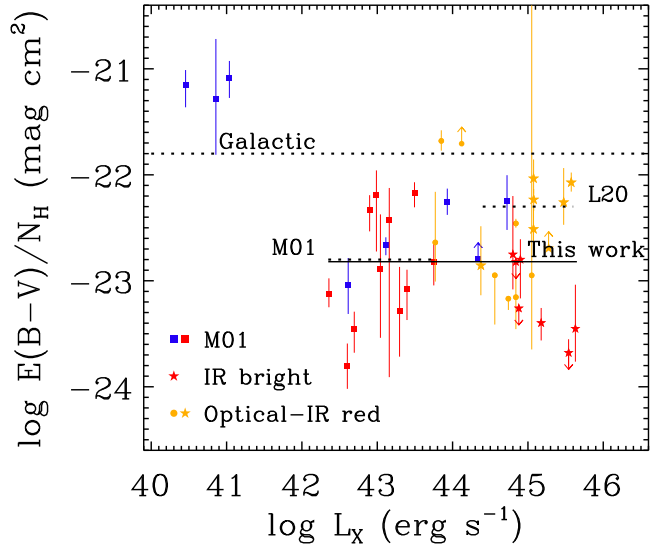}
\caption{Log ratio between $E(B-V)_{\rm cont/bl}$ and $N_{\rm H}$, plotted against intrinsic 2--10 keV luminosity. We plot only the data with $N_{\rm H}$ uncertainty values based on sufficient X-ray counts. The data come from \citet{Mai01} (marked M01, blue and red squares for type 1--1.5 and 1.8--2 classifications, respectively), red quasars (L16; G17; L17, yellow circles), heavily reddened quasars (L20, yellow stars), and Hot DOGs (S14; A16; R17a; Z18; A20, red stars). We show Galactic, L20, and \citet{Mai01} $\log E(B-V)/N_{\rm H}$ values and the range of their applicable luminosities for obscured AGNs (dashed line), and likewise from this work (solid line).}
\end{figure}

The $N_{\rm H}$ and $E(B-V)$ values are thought to be nuclear obscuration close to the AGN center, but as the AGN geometry consists of an extended, kpc order narrow line region outside the central dusty structure (e.g., \citealt{Kan18}; \citealt{Min19}), we expect the narrow-line-based $E(B-V)_{\rm nl}$ values to be smaller than the $E(B-V)_{\rm cont/bl}$ values measured closer to the nucleus (e.g., \citealt{Zak03}; \citealt{Zak05}; \citealt{Gre14}; \citealt{Jun20}). For the R17c sample providing narrow Balmer decrements, we find that the $E(B-V)_{\rm nl}/N_{\rm H}$ values for obscured (type 1.8--2.0) AGNs are about an order of magnitude smaller than the $E(B-V)_{\rm cont}/N_{\rm H}$ values in Figure 3, although showing an even larger scatter. This demonstrates that $E(B-V)_{\rm nl}$ is simply much lower than $E(B-V)_{\rm cont}$. Furthermore, the dust-to-gas ratios may decrease when using the global $N_{\rm H}$, as it is larger than the nuclear line-of-sight $N_{\rm H}$, such as for red quasars in L16, implying extended gas. The extended obscuration in obscured quasars will be considered to assess the effect of radiation pressure (\S6.2), but for a better comparison of nuclear dust and gas obscuration, we remove $E(B-V)_{\rm nl}$ values from further analysis, and we use $E(B-V)_{\rm cont}$ as the fiducial estimate of $E(B-V)$ hereafter.

\section{Eddington Ratio}
Estimating $f_{\rm Edd}$ throughout the samples relies on several bolometric correction methods and black hole scaling relations. For the bolometric correction, we primarily relied on the hard X-ray (2--10\,keV) luminosity to minimize the absorption correction for the X-ray samples in Table 1. The 2--10 keV intrinsic luminosities are based on a simple conversion of the 14--195 keV luminosities from R17c using a typical X-ray spectral slope; namely, $L_{\rm 2-10 keV} = 2.67 L_{\rm 14-195 keV}$ \citep{Rig09}. The absorption-corrected X-ray-to-bolometric correction depends on $L_{\rm bol}$ or $f_{\rm Edd}$ (e.g., \citealt{Marc04}; \citealt{Vas07}; \citealt{Lus12}). We used the \citet{Marc04} bolometric correction as a function of $L_{\rm bol}$, as the dynamic range of $L_{\rm bol}$ ($\sim$3--4 dex) is wider than that of $f_{\rm Edd}$ ($\sim$2 dex). When the X-ray luminosity was absent, we adopted the monochromatic bolometric correction from IR or extinction-corrected UV/optical continuum or line luminosities, which are relatively insensitive to $L_{\rm bol}$ (e.g., \citealt{Ric06}; \citealt{Lus12}). We used the corrections from $L_{\rm 1350}$, $L_{\rm 3000}$, and $L_{\rm 5100}$\footnote{Throughout, subscripts of $L$ indicate monochromatic continuum luminosity at that wavelength, measured in units of \AA.} (3.81, 5.15, and 9.26, respectively, \citealt{She11}) for optical quasars (J13; V18b) and obscured AGNs with extinction-corrected continuum luminosities (B12; B13; A15; B15a; objects in L17 without $N_{\rm H}$; T19; A20), $L_{\rm P\beta}$ ($\log L_{\rm bol}/10^{44} \ergs = 1.29 +0.969 \log L_{\rm P\beta}/10^{42} \ergs$, \citealt{Kim15}) for K18, $L_{\rm 3.4\mu m}$ (8, \citealt{Ham17}) for P19, $L_{\rm 15\mu m}$ (9, \citealt{Ric06}) for U12, with each correction having systematic uncertainties of a few tens of percent up to a factor of few (e.g., \citealt{Hec04}; \citealt{Ric06}; \citealt{Lus12}).

We estimated $M_{\rm BH}$ mostly through stellar absorption or broad emission lines, using a mutually consistent methodology. The mass constant of single-epoch estimators for AGNs ($f$-factor), is determined assuming that the reverberation mapped AGNs lie on the $M_{\rm BH}$--$\sigma_{*}$ relation for inactive galaxies.
We thus use the same $M_{\rm BH}$--$\sigma_{*}$ relation (e.g., \citealt{Woo15}),  
\begin{eqnarray}\begin{aligned}
\log &\Big(\frac{M_{\rm BH}}{M_{\odot}}\Big)=(8.34\pm0.05)\\
&+(5.04\pm0.28)\log \Big(\frac{\sigma_{*}}{200\kms}\Big),
\end{aligned}\end{eqnarray}
to estimate $\sigma_{*}$-based $M_{\rm BH}$ values for narrow-line AGNs where the host absorption lines are better seen, and to derive the $f$-factor in the broad FWHM-based single-epoch $M_{\rm BH}$ estimators for broad-line AGNs where AGN emission dominates over the host galaxy. The $M_{\rm BH}$($\sigma_{*}$) estimates based on other $M_{\rm BH}$--$\sigma_{*}$ relations with a shallower slope, e.g., \citet{Kor13}, are systematicaly offset to Equation (2) by 0.35 and $-$0.05 dex at $\sigma_{*}=100$ and 400\kms, respectively.

\begin{deluxetable*}{ccccc}
\tablecolumns{5}
\tabletypesize{\scriptsize}
\tablecaption{AGN $M_{\rm BH}$ estimators}
\tablehead{
\colhead{Type} & \colhead{$a$} & \colhead{$b$} & \colhead{$c$} & \colhead{$d$}}
\startdata
$M_{\rm BH}$($L_{1350}$, FWHM$_{\rm C\,IV}$) & 6.99$\pm$0.16 & 0.547$\pm$0.037 & 2.11$\pm$0.11 & 0\\
$M_{\rm BH}$($L_{1350}$, FWHM$_{\rm C\,IV}$, $\Delta \rm v_{C\,IV}$) & 6.62$\pm$0.16 & 0.547$\pm$0.037 & 2.11$\pm$0.11 & 0.335$\pm$0.022\\
$M_{\rm BH}$($L_{3000}$, FWHM$_{\rm Mg\,II}$) & 6.57$\pm$0.13 & 0.548$\pm$0.035 & 2.45$\pm$0.06 & 0\\
$M_{\rm BH}$($L_{5100}$, FWHM$_{\rm H\beta}$) & 6.88$\pm$0.12 & 0.533$\pm$0.034 & 2 & 0\\ 
$M_{\rm BH}$($L_{5100}$, FWHM$_{\rm H\alpha}$) & 6.99$\pm$0.12 & 0.533$\pm$0.034 & 2.12$\pm$0.03 & 0\\
$M_{\rm BH}$($L_{\rm P\beta}$, FWHM$_{\rm P\beta}$) & 7.24$\pm$0.16 & 0.45$\pm$0.03 & 1.69$\pm$0.16 & 0\\ 
$M_{\rm BH}$($L_{\rm P\alpha}$, FWHM$_{\rm P\alpha}$) & 7.20$\pm$0.16 & 0.43$\pm$0.03 & 1.92$\pm$0.18 & 0\\
\enddata
\tablecomments{$L$ is the continuum or broad-line luminosity, FWHM is the full width at half maximum of the best-fit broad-line model, and $\Delta \rm v_{CIV}$ is the broad \ion{C}{4} line offset to the systemic redshift (\citealt{She11} in J13, negative for blueshifts). $a$, $b$, $c$, $d$ are the coefficients in Equation (3).} 
\end{deluxetable*} 

The single-epoch $M_{\rm BH} (L, \rm FWHM)$ estimators were empirically calibrated between H$\beta$ and H$\alpha$, \ion{Mg}{2}, or \ion{C}{4} (\citealt{Jun15} using the \citealt{Ben13} $R_{\rm BLR}$--$L$ relation) with \ion{C}{4} blueshift correction when broad-line shifts were available (\citealt{Jun17}), or were calibrated between hydrogen Balmer and Paschen series (\citealt{Kim10}\footnote{Using our adopted cosmology, we find that the $R_{\rm BLR}$ values from \citet{Ben13} are higher than from \citet{Ben09} by 0.00--0.03 dex for the luminosity range used to derive Paschen line $M_{\rm BH}$ values ($L_{5100}=10^{43.5-46}\ergs$, A08; K18). K18 also note that using a single Gaussian to fit the broad Paschen lines will underestimate the $M_{\rm BH}$ values by 0.06--0.07 dex, but these amounts are negligible compared to the significance of the results (\S5).}, using the \citealt{Ben09} $R_{\rm BLR}$--$L$ relation), over a wide range of redshift and luminosity. This approach reduces the systematic offset from the choice of emission line by up to an order of magnitude\footnote{We note that a nonlinear relation between $\sigma$ and FWHM values could further result in positive/negative biases in the FWHM-based $M_{\rm BH}$ estimate at notably high and low FWHM values (e.g., \citealt{Pet04}; \citealt{Col06}), as well as whether to construct the UV or IR mass estimators to match the $M_{\rm BH}$ values to the Balmer line based, or to match the UV or IR broad-line widths and the luminosities to the optical values separately. Our choice of $M_{\rm BH}$ estimators has its own merits and limitations, and we test the systematic uncertainty of $M_{\rm BH}$ in \S5.} at extreme $M_{\rm BH}$ values \citep{Jun15}, or at extreme \ion{C}{4} blueshifts \citep{Jun17}. The estimators were updated using a common $f$-factor and uncertainty of $1.12 \pm 0.31$ for the FWHM-based $M_{\rm BH}$ \citep{Woo15}, as shown below:
\begin{eqnarray}\begin{aligned}
\log \Big(&\frac{M_{\rm BH}}{M_{\odot}}\Big)=a+\log\Big(\frac{f}{1.12}\Big)+b\,\log\Big(\frac{L}{\rm 10^{44} \ergs}\Big)\\
&+c\,\log\Big(\frac{\mathrm{FWHM}}{\rm10^{3}\kms}\Big)+d\,\log\Big(\frac{\Delta \mathrm{v_{C\,IV}}}{\rm10^{3}\kms}\Big).
\end{aligned}\end{eqnarray}
The set of $(a, b, c, d)$ values for the combination of $M_{\rm BH}$($L$, FWHM, $\Delta \rm v$) are given in Table 2. For broad-line AGNs with X-ray observations and single-epoch UV/optical $M_{\rm BH}$ estimates, we converted the X-ray-based $L_{\rm bol}$ into $L_{1350}$, $L_{3000}$, $L_{5100}$ using the aforementioned bolometric corrections, to minimize host galaxy contamination in 1350--5100\AA. We removed sources with Balmer line widths similar to [\ion{O}{3}] (A08) to distinguish broad lines from broadening by ionized gas outflows. We also limited the FWHM values to $\le$10,000\kms\ where values otherwise (e.g., 4\%\ of the J13 sample) are potentially affected by rotating accretion disks and show double-peaked lines (e.g., \citealt{Che89}; \citealt{Era94}; Table 4 in \citealt{Jun17}). Meanwhile, R17c removed single-epoch $M_{\rm BH}$ estimates for $N_{\rm H}\ge10^{22}\, \rm cm^{-2}$ AGNs as the emission line profiles could be modified by obscuration or are dominated by the narrow component \citep{Kos17}.  However, as we already removed type $\ge$1.8 sources when estimating $M_{\rm BH}$ for broad-line AGNs, we keep the $N_{\rm H}\ge10^{22}\, \rm cm^{-2}$ sources. These obscured type $\le$1.5 AGNs with $M_{\rm BH}$(FWHM) in R17c do not significantly change the distribution of $N_{\rm H}$--$f_{\rm Edd}$ with respect to using $M_{\rm BH}$($\sigma_{*}$) values. This hints that obscuration does not significantly bias the single-epoch $M_{\rm BH}$ estimates for broad-line AGNs, also consistent with the independence of broad \ion{C}{4}-to-H$\beta$ line width ratios with respect to the continuum slope for type 1 quasars (e.g., \citealt{Jun17}). We thus carefully selected only the type $\le$1.5 sources when using rest-frame UV--optical spectra to estimate $M_{\rm BH}$(FWHM) for AGNs.

Among single-epoch $M_{\rm BH}$ estimates with multiple broad-line detections, we adopted the estimators in the order of decreasing rest wavelength, while direct dynamical (B15b; B16; R17c) or reverberation-mapped (\citealt{Ben15} in R17c) $M_{\rm BH}$ values were adopted over other methods. Hot DOGs, which are heavily obscured AGNs typically showing strong, narrow lines, often display signatures of narrow-line outflows instead of ordinary broad emission lines (e.g., \citealt{Wu18}; \citealt{Jun20}). Unless the sources are thought to show scattered or leaked light from the broad-line region (A16; A20), we utilized the SED fit from A15 when deriving the $M_{\rm BH}$ constraints. Applying their maximal stellar mass ($M_{*}$) estimates from the SED fit, we gave upper limits to the $M_{\rm BH}$ values using the $M_{\rm BH}$--$M_{*}$ relation. The $M_{\rm BH}/M_{*}$ values are thought to evolve less with redshift ($\propto (1+z)^{\gamma}$, $\gamma\lesssim1$) than $M_{\rm BH}/M_{\rm bulge}$ (e.g., \citealt{Ben11}; \citealt{Din20}; \citealt{Suh20}). We adopt $M_{\rm BH}/M_{*}\sim0.003$ from the $z\sim$1--2 AGNs in \citet{Din20} and \citet{Suh20}. The same relation was used to estimate $M_{\rm BH}$ for the DOGs in C16 and T20. 

Although this analysis attempted to consistently estimate $M_{\rm BH}$ for the various samples, systematic uncertainties of a factor of several are expected from the intrinsic scatter in the BH--host mass scaling relations (e.g., \citealt{Kor13}) and the $R_{\rm BLR}$--$L$ relation (e.g., \citealt{Ben13}; \citealt{Du14}). Overall, the compiled $L_{\rm bol}$ and $M_{\rm BH}$ estimates each have systematic uncertainties of up to a factor of several or more, and although the AGN $M_{\rm BH}$ estimators include the $\sim L^{0.5}$ dependence, reducing uncertainty from the bolometric correction going into $f_{\rm Edd}\propto L_{\rm bol}/M_{\rm BH}$, we still expect systematic uncertainties of a factor of several in $f_{\rm Edd}$. We thus will interpret only the group behavior of each AGN sample within the uncertainties in $f_{\rm Edd}$.

\section{Results}
\begin{figure*}
\centering
\includegraphics[scale=0.95]{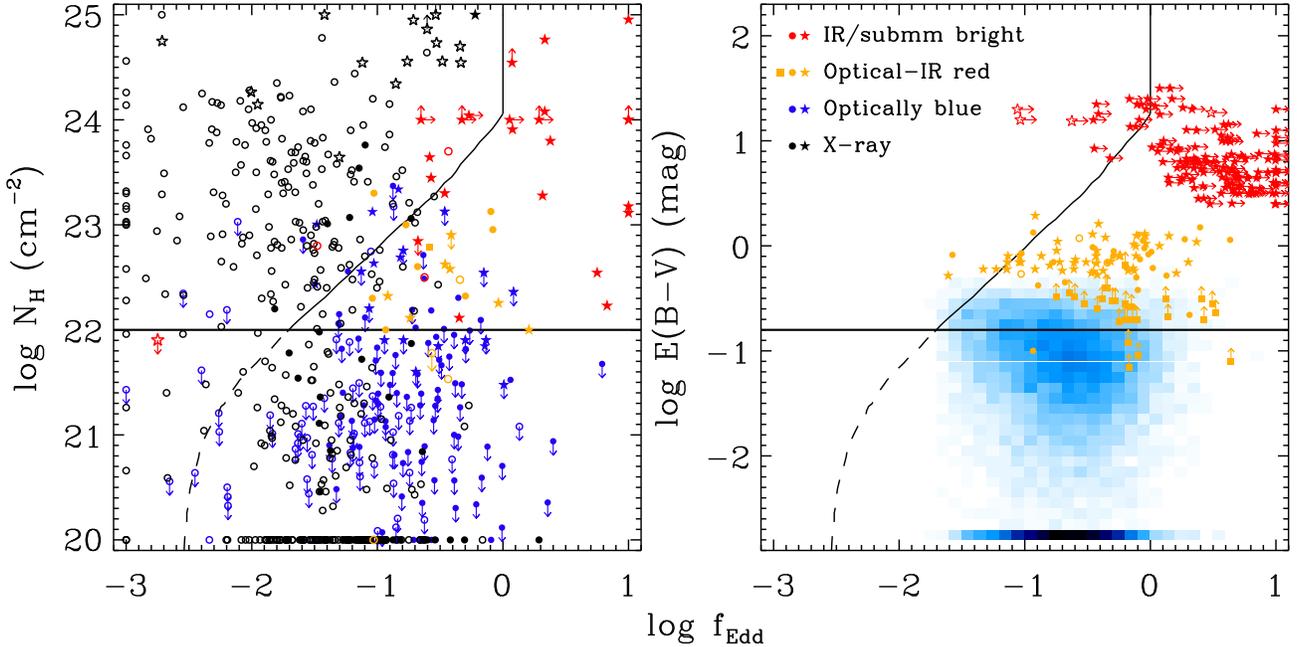}
\caption{The $N_{\rm H}$--$f_{\rm Edd}$ plane showing AGNs selected at different wavelengths. Horizontal lines separate obscured/unobscured AGNs at $N_{\rm H}=10^{22}\, \rm cm^{-2}$, and the effective Eddington ratio curves are plotted as solid/dashed lines with respect to $N_{\rm H}=10^{22}\, \rm cm^{-2}$. The symbols and color format follow that of Figure 2, except that the symbols are now filled when $L_{\rm bol} \ge 10^{45.7}\ergs$ and open otherwise. Data outside the plotted region are shown along the boundary.}
\end{figure*}
In Figure 4 we plot the distributions of $N_{\rm H}$--$f_{\rm Edd}$ and $E(B-V)$--$f_{\rm Edd}$ for the collection of AGN samples. It is clear that the forbidden region for dusty gas (Figure 1), previously less occupied by the AGNs from R17c, is well populated with IR/submillimeter-selected and optical--IR red quasars, with a minor fraction of type 1 quasars. This is seen in both the $N_{\rm H}$--$f_{\rm Edd}$ and the $E(B-V)$--$f_{\rm Edd}$ diagrams. We investigate this further in Figure 5 where we show the fraction of sources in the forbidden zone (i.e., $N_{\rm H} \ge 10^{22}\, \rm cm^{-2}$, $f_{\rm Edd, dust}>1$) per bolometric luminosity bin as a function of $L_{\rm bol}$. In the following, this fraction is referred to as $\varphi$. It is clear that $\varphi$ is minimal among the X-ray-selected AGNs with $\varphi_{N_{\rm H}}\lesssim$10\%\ at $L_{\rm bol} \sim 10^{42-47}\ergs$. Similarly, optically selected quasars have $\varphi_{N_{\rm H}}$ and $\varphi_{E(B-V)}\lesssim$20\%\ at $L_{\rm bol} \sim 10^{44-48}\ergs$, with some uncertainty for $\varphi_{N_{\rm H}}$ at $L_{\rm bol} \sim 10^{47-48}\ergs$. In contrast, the optical--IR red and IR/submillimeter-bright quasars (hereafter referred to together as luminous, obscured quasars) commonly lie mostly in the forbidden region over a wide range of $N_{\rm H}$ and $E(B-V)$ values, and we combined their statistics.\footnote{A potential caveat is the difference in the dust-to-gas ratio observed between optical--IR red and IR/submillimeter-bright quasars, which may bias the $\varphi$ value between the populations. The average dust-to-gas ratios for each population from \S3, are $\langle\log E(B-V)/N_{\rm H}\rangle=-22.33$ and $-22.88$, respectively. Using the separate ratios, however, $\varphi_{E(B-V)}$ (Figure 5 right) still remains consistent between the two populations.} The luminous, obscured quasars show $\varphi_{N_{\rm H}}$ and $\varphi_{E(B-V)}\gtrsim$\,60\%\ at $L_{\rm bol} \sim 10^{46-48}\ergs$, significantly higher than the less-luminous X-ray AGNs at comparable obscuration, or the comparably luminous but less obscured optical quasars. These findings confirm earlier studies by \citet{Gli17b} and L20 on optical--IR red quasars, with our results applicable to general luminous, obscured quasars.

Systematic uncertainties of a factor of several in $f_{\rm Edd}$ (\S4) may change the fraction of the samples in the forbidden region. We test this by giving a $\pm$0.5 dex offset to the $f_{\rm Edd, dust} (N_{\rm H})=1$ curve and recalculating $\varphi$. The $\varphi$ values are nearly unchanged for the X-ray AGNs and optical quasars, whereas for the luminous, obscured quasars, $\varphi$ may drop down to 40\%--50\% at $L_{\rm bol} \sim 10^{46-48}\ergs$ if the observed $f_{\rm Edd}$ values are overestimated by 0.5 dex. Still, the $\varphi$ values for the luminous, obscured quasars are several times the X-ray AGNs or optical quasars at a given luminosity, and the main trend in Figure 5 remains unchanged. Modifications to the $f_{\rm Edd, dust}$ curve may also occur when considering the effect of dust-to-gas ratios closer to the Milky Way value than the value adopted in this work, or radiation trapping. The enhanced absorption of the incident radiation by dust or trapping of reprocessed radiation lowers the $f_{\rm Edd, dust}$ curve, at $N_{\rm H} \ge 10^{22}\, \rm cm^{-2}$ \citep{Ish18}. Still, we note that both effects simply increase $\varphi$ for luminous, obscured quasars, reinforcing our findings in Figures 4 and 5.

The $f_{\rm Edd, dust}$ values can be further shifted by nuclear stars.
Adopting the sphere of influence from the BH, we have
\begin{eqnarray}\begin{aligned}
r_{\rm BH}=GM_{\rm BH}&/\sigma_{*}^{2}\\
=107\,{\rm pc}\,&\Big(\frac{M_{\rm BH}}{10^{9}\,M_{\odot}}\Big)\Big(\frac{\sigma_{*}}{200\,\kms}\Big)^{-2},
\end{aligned}\end{eqnarray} 
and the enclosed mass $M(<r)$ becomes multiple times larger than $M_{\rm BH}$ at $r \gtrsim r_{\rm BH}$ due to nuclear stars, increasing the $L_{\rm Edd, dust}$ in Equation (1) for a radiation pressure force to balance off a stronger gravity by $M(<r)/M_{\rm BH}$ times. Nuclear stars may lower $f_{\rm Edd, dust}$ significantly if the BH is undermassive relative to the $M_{\rm BH}$--$M_{*}$ relation, or if the stellar light is more concentrated near the center. To fully explain the excess $\varphi$ in luminous, obscured quasars by the shift in $L_{\rm Edd, dust}$ values, however, their negative offset on the $M_{\rm BH}$--$M_{*}$ relation should be on average an order of magnitude, which is less likely from current observations (e.g., \citealt{Bor14}, but see also U12, where a fraction of their sample show undermassive $M_{\rm BH}$/$M_{*}$ values). Instead, assuming $M_{\rm BH}/M_{*}\sim0.003$ from $z\sim$1--2 AGNs (\citealt{Din20}; \citealt{Suh20}), we find that the S{\'e}rsic index \citep{Ser63} for luminous, obscured quasars should still be $\approx$2.5--3 times larger than for the typical AGN for $M(<r_{\rm BH})/M_{\rm BH}$ to be boosted by an order of magnitude, which is likewise a stringent condition. Therefore, we find that nuclear stars themselves are less likely to shift $f_{\rm Edd, dust}$ values enough to falsify our results.

We note that the IR-luminous, obscured quasars appear to show higher $\varphi_{N_{\rm H}}$ values compared to obscured (type 1.8--2), X-ray-selected quasars at matched luminosity. The type 1.8--2 sources in R17c have $\varphi_{N_{\rm H}}$ values $\lesssim$20\%, still avoiding the forbidden region by factors of several compared to the IR-luminous, obscured quasars at $L_{\rm bol} \gtrsim 10^{46}\ergs$. We caution that the IR luminous, obscured quasars are intrinsically different from the obscured, BAT-selected quasars, however, as the $\it Swift$/BAT survey is missing the distant, obscured quasar population often detected by deeper X-ray observations (e.g., \citealt{Mer14}; see the references in Table 1). 

\begin{figure*}
\centering
\includegraphics[scale=0.95]{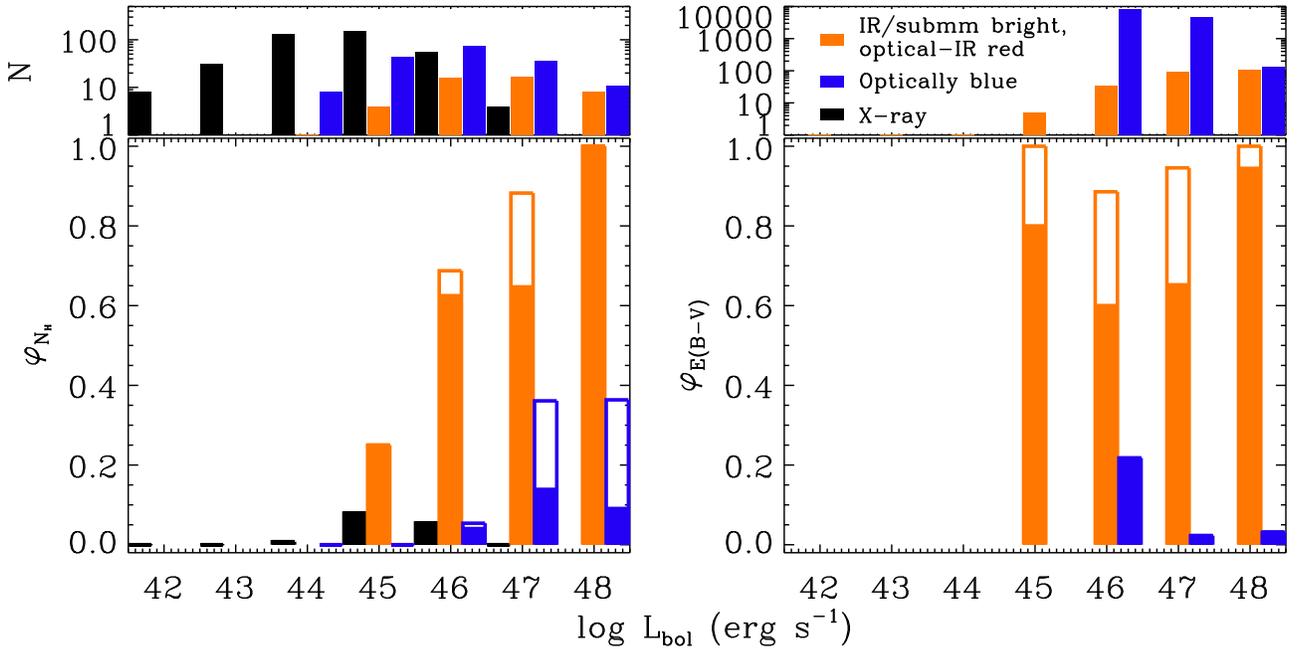}
\caption{Fraction of AGNs in the forbidden region ($\varphi$) on the $N_{\rm H}$--$f_{\rm Edd}$ (bottom left) and $E(B-V)$--$f_{\rm Edd}$ (bottom right) planes, with the number of sources used to plot the fractions in each luminosity bin (top). The solid graphs indicate the minimum value of $\varphi$ for the given upper and lower limits on obscuration or $f_{\rm Edd}$, whereas the open graphs include all of the upper and lower limits enclosing the forbidden region. The colors for X-ray and optically blue AGNs follow those of Figures 2 and 4, whereas we merge the optical--IR red and IR/submillimeter-bright samples that show similar $\varphi$ values, in orange. The graphs are plotted when the number of AGNs $\ge4$ per bin, offset to each other only for display purposes, and are calculated based on a common interval in $L_{\rm bol}$.}
\end{figure*}

\section{Discussion}
The luminous, obscured quasars are known to be a substantial population in the intermediate-redshift universe, as their combined number density is comparable to that of unobscured quasars at matched luminosity (A15; B15a; \citealt{Ham17}; \citealt{Gli18}). As the typical $M_{\rm BH}$ values for $L_{\rm bol}>10^{46} \ergs$ quasars in this work are similar ($\sim10^{9}M_{\odot}$) for both the obscured and unobscured populations, the timescale of the luminous, obscured quasar phase with $f_{\rm Edd, dust}>1$ is presumed to be similar to that of the $f_{\rm Edd, dust}<1$ quasars. In contrast, the nearly complete absence of lower-luminosity AGNs with $f_{\rm Edd, dust}>1$ (\S5) suggests a much shorter obscured phase for lower-luminosity AGNs. This appears as a challenge for the radiation-pressure feedback in regulating the nuclear obscuration for luminous quasars, and we next consider possible evolutionary scenarios to achieve a coherent picture of dust obscuration in luminous quasars. 

\subsection{Active timescale}
First, the AGN timescale (hereafter $t_{\rm AGN}$) is thought to be shorter for more luminous quasars, and this may explain higher $\varphi$ values for luminous quasars. The nearby AGN fraction is measured to be tens of percent of the galaxy lifetime (e.g., \citealt{Ho97}; \citealt{Kau03}), with a corresponding $t_{\rm AGN}$ of $\sim10^{9}$\,yr assuming typical galaxy lifetimes of $\sim10^{10}$\,yr. More luminous quasars are more rare, with expected $t_{\rm QSO}\sim10^{7-8}$\,yr (e.g., \citealt{Mart04}; \citealt{Hop05}; \citealt{Hop09}). To explain $\varphi\lesssim$1--10\% for $L_{\rm bol} \lesssim 10^{44}\ergs$ AGNs in Figure 5, we constrain the timescale for radiation pressure to clear the nuclear obscuration (hereafter the radiation feedback timescale, $t_{\rm rad}=t_{\rm AGN}\,\varphi$) to be $t_{\rm rad}\lesssim10^{9}\,(0.01-0.1)\sim10^{7-8}$\,yr. Assuming that luminous, obscured quasars will evolve into comparably luminous unobscured quasars through radiation-pressure feedback, so as to explain the comparable number density between the populations, $t_{\rm rad}$ for luminous ($L_{\rm bol} \gtrsim 10^{46}\ergs$) quasars with $\varphi\gtrsim$\,60\%\ would be $t_{\rm rad}\sim0.5t_{\rm QSO}\,\varphi \sim0.5(10^{7-8})(0.6-1)=(3-5)\times10^{6-7}$\,yr, roughly comparable to $t_{\rm rad}$ for less-luminous AGNs. We note that if AGN activity is more episodic (e.g., \citealt{Par11}; \citealt{Yaj17}), feedback timescales may be shortened accordingly, although it seems more likely that luminous quasars have few episodes of vigorous accretion (e.g., \citealt{Hop09}). Luminous, obscured quasars may thus appear to show higher $\varphi$ values due to a shorter $t_{\rm AGN}$ than less-luminous, obscured AGNs, even if they feel the same radiation pressure.

We have referred to $t_{\rm QSO}\lesssim10^{7-8}$\,yr for quasars as a whole (e.g., $M_{\rm B}\lesssim-23$ mag or $L_{\rm bol}\gtrsim10^{45}\ergs$), but if more luminous, obscured quasars are in a shorter phase of AGN evolution (shorter $t_{\rm AGN}$), it better explains the highest $\varphi$ values observed at $L_{\rm bol} \gtrsim 10^{46}\ergs$. L20 note outflow timescales ($t_{\rm out}$) for nuclear obscuration to clear away in an expanding shell by radiation pressure on dust,
\begin{equation}
t_{\rm out} \approx 2\times10^{5}\,{\rm yr}\,\Big(\frac{r_0}{30\rm pc}\Big) \Big(\frac{v_{\rm out}}{1000\kms}\Big)^{-1}
\end{equation}
finding $t_{\rm out} \approx 2\times 10^{5}$ yr for Compton-thick gas expanding from an initial distance of $r_{0}=$30 pc until it reaches $N_{\rm H}=10^{22}\, \rm cm^{-2}$, assuming $v_{\rm out}=10^{3}$\kms. If the dusty gas outflows are triggered by radiation pressure, we expect $t_{\rm out}$ to be equal to $t_{\rm rad}$. However, it is shorter than our estimated $t_{\rm rad}$ values for luminous quasars, by $\sim \{(3-5)\times10^{6-7}\}/(2\times 10^{5})$ or $\sim$1--2 orders of magnitude. This can be explained if $t_{\rm AGN}$ for $L_{\rm bol}\gtrsim10^{46}\ergs$ quasars are $\sim$1--2 orders of magnitude shorter than the $t_{\rm QSO}\sim10^{7-8}$\,yr we adopt, qualitatively consistent with the drop of $t_{\rm AGN}$ for more luminous quasars in simulations (e.g., \citealt{Hop05,Hop06}).

\subsection{Extended obscuration}
An alternative description is that it takes a longer $t_{\rm rad}$ for luminous, obscured quasars to clear their obscuration than at lower luminosity. Radiation pressure from luminous, obscured quasars should effectively reach larger distances in the galaxy according to the decreasing small-scale dust covering factor observed in high $L_{\rm bol}$ or $f_{\rm Edd}$ AGNs (e.g., \citealt{Mai07}; \citealt{Tob14}; \citealt{Ezh17}). Thus, observing high $\varphi$ values in luminous, obscured quasars implies that dusty gas may be spatially extended into their hosts, in contrast to lower-luminosity AGNs. This is supported by observations of obscured quasars showing an extended distribution of disturbed emission (e.g., \citealt{San88}; \citealt{San96}; \citealt{Urr08}; \citealt{Gli15}; \citealt{Fan16}; \citealt{Ric17b}). An increased fraction of obscured yet broad-line AGNs (e.g., \citealt{Lac15}; \citealt{Gli18}) or extended dust extinction through lines of sight kiloparsecs away from narrow-line AGNs are seen (\S3) at quasar luminosities, also in agreement with global column densities much larger than the line-of-sight $N_{\rm H}$ for red quasars (\S3).
\begin{figure}
\centering
\includegraphics[scale=0.95]{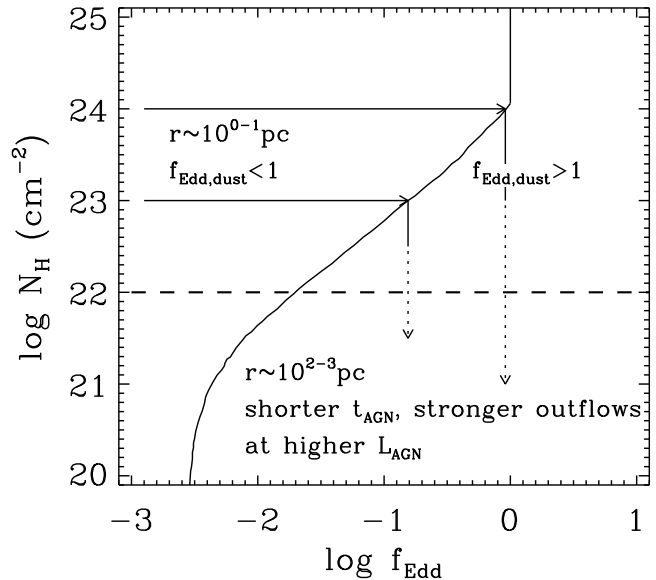}
\caption{A schematic diagram showing the maximally allowed accretion rate and radiation pressure/outflow track on the $N_{\rm H}$--$f_{\rm Edd}$ plane for $N_{\rm H}=10^{23}, 10^{24}\, \rm cm^{-2}$ quasars. At $L_{\rm bol}\lesssim10^{45}$\ergs, the lack of objects in the forbidden region suggests that radiation pressure on dusty gas controls nuclear ($\lesssim$10 pc order) obscuration and quickly drops the obscuration down to $N_{\rm H}\lesssim10^{22}\, \rm cm^{-2}$, while extended AGN outflows are less observed. At $L_{\rm bol}\gtrsim10^{46}$\ergs, $f_{\rm Edd, dust}>1$ accretion (solid line) may likewise clear the nuclear obscuration, but the high fraction of obscured quasars in the forbidden region suggests a short luminous quasar timescale and/or an extended, $\sim10^{2-3}$ pc-scale obscuration being cleared slowly by outflows (dotted line).}
\end{figure}

According to simplified models for AGNs and host galaxy obscuration at multiple scales (e.g., \citealt{Buc17}; \citealt{Hic18}), the host galaxy kiloparsecs away from the nucleus is responsible for $N_{\rm H} \sim 10^{22}\, \rm cm^{-2}$, whereas obscuration from the inner AGN structure ($\lesssim$10\,pc) or circumnuclear starbursts ($\sim$10--100\,pc) can reach Compton-thick column densities. In addition, obscuration from gas-rich mergers (e.g., \citealt{Hop08}) or higher gas fractions in high-redshift galaxies (e.g., \citealt{Tac10}; \citealt{Buc17}) may enhance the obscuration up to kiloparsec scales. Coming back to Equation (5), we find that $t_{\rm out}$ for luminous, obscured quasars will be extended by 1--2 orders of magnitude ($t_{\rm out}\sim10^{6-7}$ yr) if extended obscuration due to mergers is spread over $\sim10^{2-3}$ pc, closing the gap between the timescale arguments (\S6.1) without even changing $t_{\rm AGN}$. This scenario is also consistent with the $t_{\rm out}$ values estimated by modeling expanding shells of dusty gas located $\sim10^{2-3}$ pc away from luminous quasars \citep{Ish17}. We direct readers to the theoretical and observational studies on how the nuclear outflows triggered by radiation pressure extend to the host galaxy (e.g., \citealt{Har14}; \citealt{Ish15}; \citealt{Tho15}; \citealt{Ish17}; \citealt{Kan18}).

Although the impact of radiation pressure from the AGN itself is weaker at extended regions of the galaxy, and R17c separate radiation-pressure feedback from inflows and outflows, radiation pressure has still been considered to launch outflows that may reach large distances (e.g., \citealt{Hop10} and the discussion in L20). In this work, we separately considered radiation pressure to regulate $\lesssim10$ pc order obscuration and outflows $\sim10^{2-3}$pc scales, but note that radiation pressure is thought to cause extended outflows that eventually clear obscured quasars, according to gas-rich, merger-driven quasar evolution models (e.g., \citealt{Hop08}; \citealt{Hic09}). Not only are the highly ionized gas outflows on the order of $\sim10^{3}\kms$ found in the majority of quasars with $L_{[\rm O\,III]}\gtrsim10^{42}$\ergs\ (or $L_{\rm bol}\gtrsim10^{45.5}$\ergs), or $f_{\rm Edd} \gtrsim 0.1$ (e.g., \citealt{Woo16}; \citealt{Rak18}; \citealt{Shi19}; \citealt{Jun20}), they extend over kiloparsec scales together with Balmer line outflows with a weaker ionization potential or molecular outflows (e.g., \citealt{Fio17}; \citealt{Kan17}; \citealt{Fle19}). This is in line with higher merger fractions seen in $L_{\rm bol}\gtrsim10^{46}$\ergs\ quasars (e.g., \citealt{Tre12}; \citealt{Fan16}; \citealt{Dia18}), which is also the transitional luminosity where radiation-pressure feedback appears less effective (Figure 5). 

We thus consider radiation pressure to be responsible for regulating not only the $\lesssim$10 pc-order dusty structure (e.g., \citealt{Law91}; R17c), but also the host galaxy environment in obscured $L_{\rm bol} \gtrsim 10^{46}$\ergs\ quasars where the triggered nuclear outflows may reach and clear $\sim10^{2-3}$ parsec-scale material, slowly over a timescale of $\sim10^{6-7}$\,yr. This is consistent with the high-$f_{\rm Edd}$ AGN outflows discussed in R17c, though their sample lacked the luminous quasars that we argue are responsible for producing extended outflows at $f_{\rm Edd, dust}>1$ values. We summarize our discussion in Figure 6.\\

\section{Summary}
Using a collection of AGN samples spanning a wide dynamic range of luminosity, obscuration, and redshift, we probed the distribution of obscuration and accretion rate values to comparatively examine the role of radiation pressure in blowing out obscured quasars. We summarize our findings below:
 
1. The fraction of AGNs in the forbidden zone for radiation pressure, $\varphi$, is kept to $\lesssim$20\%\ for all of the multi-wavelength-selected AGN samples compiled over a wide range of luminosity and redshift, consistent with previous findings that nuclear obscuration is quickly blown away by radiation pressure once the accretion rate exceeds the Eddington limit for dusty gas.

2. This radiation-pressure feedback, that is, the acceleration of nuclear dusty gas, appears limited for luminous, obscured quasars at $N_{\rm H}\gtrsim10^{22}\, \rm cm^{-2}$ or $E(B-V) \gtrsim 0.2$ mag, and $L_{\rm bol}\gtrsim10^{46}$\ergs, where they show $\varphi\gtrsim60\%$ over a wide range of AGN selection wavelengths or amount of obscuration. This may be explained by a combination of shorter luminous quasar lifetimes and extended obscuration cleared by outflows over a longer timescale than to clear the nuclear obscuration. 

Ultimately, we expect to see the $M_{\rm BH}$ values grow while luminous, obscured quasars become unobscured if extended outflows, slower than radiation pressure clearing the nuclear obscuration, are the bottleneck for AGN feedback. Ongoing hard X-ray surveys probing fainter sources (e.g., \citealt{Lan17}; \citealt{Oh18}) will confirm if distant, gas-obscured quasars are going through similar strengths of radiation-pressure feedback as dust-obscured quasars. Spatially resolved or global $N_{\rm H}$ and $E(B-V)$ estimates for luminous, obscured quasars will better tell whether obscuration is indeed more extended in luminous quasars and will quantify the relative effect of radiation pressure and outflows to their parsec-to-kiloparsec scale gas and dust environments. 

\acknowledgments
We thank the anonymous referee for the comments that greatly improved the paper and Andrew Fabian for kindly providing the $f_{\rm Edd, dust}(N_{\rm H})=1$ curves.
This research was supported by the Basic Science Research Program through the National Research Foundation of Korea (NRF) funded by the Ministry of Education (NRF-2017R1A6A3A04005158). R.J.A. was supported by FONDECYT grant No. 1191124. R.C.H. and C.M.C. acknowledge support from the National Science Foundation under CAREER award no. 1554584. C.R. acknowledges support from the Fondecyt Iniciacion grant 11190831. 

This work makes use of data from the $\it NuSTAR$ mission, a project led by Caltech, managed by the Jet Propulsion Laboratory, and funded by NASA.
This research has made use of data and/or software provided by the High Energy Astrophysics Science Archive Research Center (HEASARC), which is a service of the Astrophysics Science Division at NASA/GSFC and the High Energy Astrophysics Division of the Smithsonian Astrophysical Observatory.
This publication makes use of data products from the United Kingdom Infrared Deep Sky Survey. UKIRT is owned by the University of Hawaii (UH) and operated by the UH Institute for Astronomy; operations are enabled through the cooperation of the East Asian Observatory. When the data reported here were acquired, UKIRT was operated by the Joint Astronomy Centre on behalf of the Science and Technology Facilities Council of the U.K.
This publication makes use of data products from the Wide-field Infrared Survey Explorer, 
which is a joint project of the University of California, Los Angeles, and the Jet Propulsion 
Laboratory/California Institute of Technology, funded by the National Aeronautics and Space Administration.

\end{document}